\documentclass[reprint,prx,aps,amsmath,amssymb,floatfix,superscriptaddress,longbibliography,cite]{revtex4-2}
 
\usepackage{mathtools,amsmath}
\usepackage{graphicx} 
\usepackage{tikz}
\usepackage[breaklinks=true,colorlinks,citecolor=teal,linkcolor=teal,urlcolor=teal]{hyperref}
\usepackage{physics}
\usepackage{siunitx}
\usepackage{bbold}
\usepackage{soul}
\usepackage{bm}
\usepackage[normalem]{ulem}
\usepackage{times}
\usepackage{enumerate}  
\usepackage{float}
\usepackage{amssymb}
\usepackage{xcolor}

\newcommand{\sigmaz}{\hat \sigma_{\rm z}}
\newcommand{\sigmax}{ \hat \sigma_{\rm x}}

\newcommand{\sz}{\hat \sigma_{\rm z}}
\newcommand{\sx}{\hat\sigma_{\rm x}}

\newcommand{\equasi}{\Delta \epsilon}
\newcommand{\qubitw}{w_{\rm q}}

\begin{document}

\title{Dynamical Sweet and Sour Regions in Bichromatically Driven Floquet Qubits}
\author{D. Dominic Brise\~{n}o-Colunga}
\thanks{These authors contributed equally; correspondence should be addressed to \href{mailto:bbhandari@chapman.edu}{bbhandari@chapman.edu}.}
\affiliation{Institute for Quantum Studies, Chapman University, Orange, CA 92866, USA}
\author{Bibek Bhandari}
\thanks{These authors contributed equally; correspondence should be addressed to \href{mailto:bbhandari@chapman.edu}{bbhandari@chapman.edu}.}
\affiliation{Institute for Quantum Studies, Chapman University, Orange, CA 92866, USA}
\author{Debmalya Das}
\affiliation{Dipartimento di Fisica, Universit\`a  di Bari, I-70126 Bari, Italy}
\affiliation{INFN, Sezione di Bari, I-70125 Bari, Italy}
\author{Long B. Nguyen}
\affiliation{Quantum Nanoelectronics Laboratory, Department of Physics, University of California, Berkeley, CA, 94720, USA}
\affiliation{Applied Mathematics and Computational Research Division, Lawrence Berkeley National Laboratory, Berkeley, CA, 94720, USA}
\author{Yosep Kim}
\affiliation{Department of Physics, Korea University, Seoul 02841, Korea}
\author{David I. Santiago}
\affiliation{Quantum Nanoelectronics Laboratory, Department of Physics, University of California, Berkeley, CA, 94720, USA}
\affiliation{Applied Mathematics and Computational Research Division, Lawrence Berkeley National Laboratory, Berkeley, CA, 94720, USA}
\author{Irfan Siddiqi}
\affiliation{Quantum Nanoelectronics Laboratory, Department of Physics, University of California, Berkeley, CA, 94720, USA}
\affiliation{Applied Mathematics and Computational Research Division, Lawrence Berkeley National Laboratory, Berkeley, CA, 94720, USA}

\author{Andrew N. Jordan}
\affiliation{Institute for Quantum Studies, Chapman University, Orange, CA 92866, USA}
\affiliation{Schmid College of Science and Technology, Chapman University, Orange, CA, 92866, USA}
\affiliation{The Kennedy Chair in Physics, Chapman University, Orange, CA, 92866, USA}
\affiliation{Department of Physics and Astronomy, University of Rochester, Rochester, NY, 14627, USA}

\author{Justin Dressel}
\affiliation{Institute for Quantum Studies, Chapman University, Orange, CA 92866, USA}
\affiliation{Schmid College of Science and Technology, Chapman University, Orange, CA, 92866, USA}

\begin{abstract}
Modern superconducting and semiconducting quantum hardware use external charge and microwave flux drives to both tune and operate devices. However, each external drive is susceptible to low-frequency (e.g., $1/f$) noise that can drastically reduce the decoherence lifetime of the device unless the drive is placed at specific operating points that minimize the sensitivity to fluctuations. We show that operating a qubit in a driven frame using two periodic drives of distinct commensurate frequencies can have advantages over both monochromatically driven frames and static frames with constant offset drives. Employing Floquet theory, we analyze the spectral and lifetime characteristics of a two-level system under weak and strong bichromatic drives, identifying drive-parameter regions with high coherence (sweet spots) and highlighting regions where coherence is limited by additional sensitivity to noise at the drive frequencies (sour spots). We present analytical expressions for quasienergy gaps and dephasing rates, demonstrating that bichromatic driving can alleviate the trade-off between DC and AC noise robustness observed in monochromatic drives. This approach reveals continuous manifolds of doubly dynamical sweet spots, along which drive parameters can be varied without compromising coherence.  Our results motivate further study of bichromatic Floquet engineering as a powerful strategy for maintaining tunability in high-coherence quantum systems. 
\end{abstract}

\maketitle

\section{Introduction}

The pursuit of robust, high-coherence qubits has driven remarkable progress in both superconducting~\cite{scqubit_fluxonium,scqubit_grimm,scqubit_nakamura,Engineers_guide_to_scq,noise_koch,noise_nguyen,bhandari2024,hajr2024} and semiconducting quantum platforms~\cite{bechtold2015,noise_ithier,Guo_dephasing,Oliver_2013_noise_spectrum_analyser,decoherence_electrical_driving_2014,Smirnov_2003}. Superconducting qubits, while offering good controllability and scalability~\cite{blais_cqed_2021,google_willow}, are more susceptible to environmental noise and decoherence, leading to shorter coherence times~\cite{noise_koch,noise_ithier}. In contrast, semiconducting spin qubits leverage atomic-scale confinement and material isolation to achieve significantly longer coherence times~\cite{kobayashi2021engineering,burkard2023review}, though their control and scalability remain technically challenging~\cite{burkard2023review,tanttu2024assessment}. However, both platforms face a critical challenge: decoherence induced by low-frequency noise~\cite{1_over_f_noise, fluxonium_high_coherence,scqubit_fluxonium,Huang2021,noise_anton,noise_hazard,noise_ithier,noise_koch,noise_long,noise_nguyen,noise_yoshihara}. For superconducting flux-tunable qubits, ubiquitous 
$1/f$ flux noise limits their static (undriven) operation to particular flux bias choices that minimize noise sensitivity, restricting tunability~\cite{rigetti_bi,rigetti_2019,Huang2021,Hajati2024}. Similarly, semiconducting spin qubits are susceptible to low-frequency noise in either electric charge or magnetic flux (depending on the design) as well as decoherence from coupling to phonon baths~\cite{burkard2023review,Hajati2024}. This necessitates dynamic error suppression strategies such as spin-echo or dynamical decoupling~\cite{petta2005coherent,medford2012,muhonen2014storing, ezzel_dynamical_2023}.

Recent advances in Floquet engineering--the use of periodic drives to reshape a quantum system’s effective Hamiltonian--enable an intriguing approach to addressing these challenges by encoding qubit information in a rotating frame that is dynamically decoupled from the low-frequency noise ~\cite{rigetti_2019,noise_nguyen,rigetti_bi}. In superconducting circuits, replacing static flux (charge) bias with monochromatic flux (charge) drives has revealed dynamical sweet spots with enhanced noise resilience, enabling high-fidelity gates~\cite{noise_nguyen,rigetti_2019,hajr2024,qing2024,frattini2024}. Parallel developments in semiconductor qubits have exploited periodic driving to suppress charge noise via spin-locking~\cite{Huang2021,noise_ithier,Guo_dephasing,Oliver_2013_noise_spectrum_analyser,decoherence_electrical_driving_2014,Smirnov_2003} and dynamical decoupling from spin baths~\cite{bluhm2011dephasing,burkard2023review,petta2005coherent}. These successes highlight a cross-platform principle: periodic drives can help decouple qubits from low-frequency noise while allowing control. However, existing strategies in both domains face limitations as monochromatic drives restrict parameter manifolds and result in poor gate operations near the dynamical sweet spots~\cite{Huang2021}.

Bichromatic driving has recently emerged as a promising strategy for noise-resilient quantum control. Recent experiments in superconducting circuits leveraged bichromatic drives to create continuous dynamical sweet spots, enabling high-fidelity single- and two-qubit gates~\cite{rigetti_bi}. Further, in Ref.~\cite{noise_nguyen}, bichromatic driving was employed to suppress static ZZ coupling between Floquet qubits, helping realize a high-fidelity two-qubit gate. Building on these advances, we theoretically investigate how bichromatic driving reshapes the qubit-environment interaction, either suppressing or enhancing the sensitivity to noise at both low frequencies and the drive frequencies. By studying the interplay between drive-engineered spectral manifolds and environmental coupling, this work establishes design principles for decoherence mitigation in dynamical qubits.
\begin{figure}
\centering
\begin{tikzpicture}
\node[inner sep=0pt] (russell) at (0,0) 
    {\includegraphics[width=.45\textwidth]{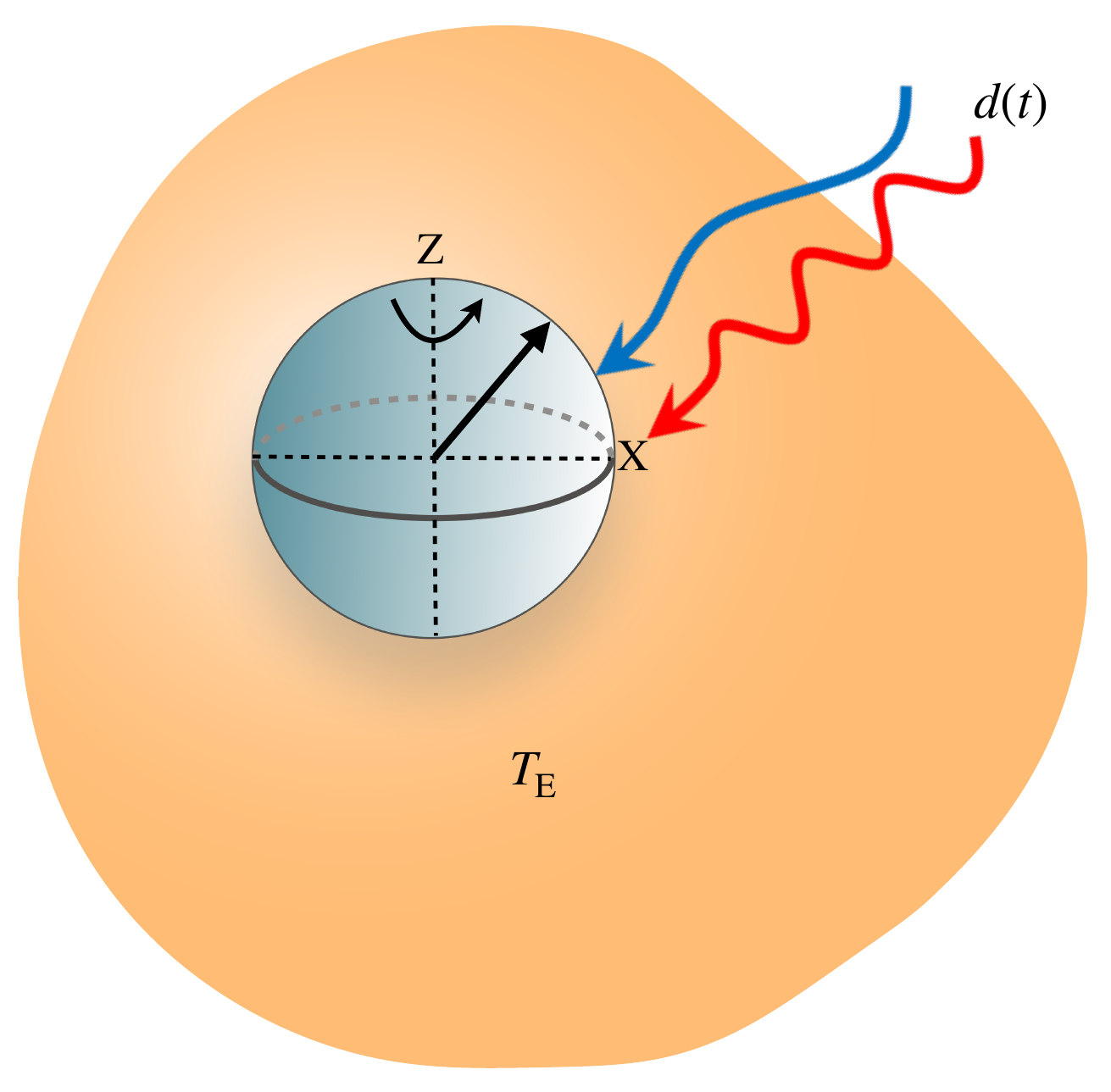}};
\node[inner sep=0pt] (russell) at (-1.2,1.4) {${{w}_{\rm q}}$};
\node[inner sep=0pt] (russell) at (0.34,1.96) {$\ket{\psi(t)}$};
\end{tikzpicture}
\caption{Schematic depiction of a two-level system (represented by a Bloch sphere) driven by a bichromatic drive $d(t) =  2\Omega \cos (\nu) \cos(N_1 \omega t) + 2\Omega \sin (\nu) \cos(N_2 \omega t) + b $, where $N_1$ and $N_2$ are integers. The qubit is linearly coupled to a thermal environment at temperature $T_{\rm E}$, leading to decoherence via $1/f$ and dielectric noise channels.}
\label{fig:sketch}
\end{figure}

In this paper, we investigate bichromatic Floquet engineering as a strategy for noise-protected qubit operation. Our analytical framework reveals universal features applicable to superconducting and semiconducting spin qubits. Using a generic two-level system as a testbed (see Sec.~\ref{sec:hamil} for the model Hamiltonian and Sec.~\ref{sec:lifetime} for the noise model), we demonstrate that bichromatic driving creates continuous high-coherence manifolds in parameter space in the weak (see Sec.~\ref{sec:weak}) and beyond weak driving regime (see Sec.~\ref{sec:beyond_weak}), where decoherence from $1/f$ flux noise is suppressed by orders of magnitude. By deriving analytic expressions for the AC Stark shift using Generalized Van Vleck (GVV) perturbation theory, we provide a detailed theoretical analysis of the resonantly and bichromatically driven qubit (see Sec.~\ref {sec:res_singl}). Our analysis directly connects the gap’s sensitivity to external drive parameters, such as modulation amplitude and drive frequency, enabling precise identification of dynamical regions of low sensitivity to DC and AC noise (doubly sweet spots).

\subsection{Driven 2-level System Hamiltonian}
\label{sec:hamil}
We consider the system Hamiltonian (setting $\hbar =1$) of the following form 
\begin{equation}
    \hat H (t) = -\frac{w_q}{2}\sz + \frac{d(t)}{2}\sx, \label{eq: hamiltonain def}
\end{equation}
where the first term on the right-hand side is the undriven qubit Hamiltonian. We define the Pauli operators as, $\sigmaz = \ket{g}\!\bra{g}-\ket{e}\!\bra{e}$ and $\sigmax = \ket{g}\!\bra{e}+\ket{e}\!\bra{g}$. We parametrize the external drive, $d(t)$, as
\begin{equation}
    d(t) = \Omega\cos(\nu)\cos(N_1\omega t) + \Omega\sin(\nu)\cos(N_2\omega t) + {b},
    \label{eq: drive def}
\end{equation}
where $b$ is the DC component of the drive, $\Omega$ sets the AC drive strength, and $\nu$ is a time-independent mixing angle. We consider commensurate drive frequencies, $N_1\omega$ and $N_2\omega$, by choosing $N_1$ and $N_2$ to be integers. We will refer to the case where $\Omega=0$ as a ``DC qubit". 

A periodically driven Hamiltonian admits orthonormal, quasi-periodic solutions called Floquet states, $\ket{\psi_\pm(t)} = e^{-i\epsilon_\pm t}\ket{u_\pm(t)}$, where $\epsilon_\pm$ is called the quasienergy, and $\ket{u_\pm(t)}$ is the $T$-periodic Floquet mode for the Floquet states $(+,-)$ (see App.~\ref{sec:floquet} for details). The quasienergies and their respective Floquet modes are eigenvalues and eigenvectors of a Hermitian operator called the Floquet Hamiltonian, $\hat H_{\rm{F}} (t)= \hat H(t) - i\frac{\partial}{\partial t}$. Further, the Floquet modes may be expanded as
\begin{equation}
    \ket {u_\pm} = \sum_{n,\alpha} c^\pm_{n,\alpha} \ket{n,\alpha}, \label{eq: time-dependent floquet modes}
\end{equation}
where $c^\pm_{n,\alpha} = \frac {1} {\rm T} \int_0^T e^{-in\omega t} \langle \alpha | u_\pm\rangle dt$ are the time-independent Floquet coefficients for two-level system state $\ket{\alpha} \equiv \{\ket{g},\ket{e}\}$. In Eq.~(\ref{eq: time-dependent floquet modes}), we set the basis for the T-periodic functions $(\mathcal T)$ as $\left \{ \ket n = e^{in\omega t} \right\}_{n\in\mathbb Z}$. Further, the state $\ket{n,\alpha}=e^{in\omega t}\ket{\alpha}$ satisfies the properties of a product space between $\mathcal T$ and atomic states of the undriven system. Changing coordinates to the extended Hilbert space determined by the basis $\ket{n,\alpha}$, the Floquet Hamiltonian can be written as $\hat H_{\rm F} = \hat H_{\rm 0} + \hat H_{\rm{DC}} +  \hat H_{\rm{AC}}$, where $\hat H_{\rm 0}$ describes the undriven qubit dynamics, $\hat H_{\rm{DC}}$ describes the action of the DC bias $b$ plus the operator $-i\partial_t$, and $\hat H_{\rm{AC}}$ describes the action of the AC drive. They are (see App.~\ref{sec:floq_ham_app})
\begin{align}
    \hat H_{\rm 0} &= \hat 1_{\mathcal T}\otimes \left(-\frac{w_q}{2}\sz\right),\label{eq: H_D}\\
    \hat H_{\rm{\rm{DC}}} &= \sum_m \ket{m}\bra{m}\otimes(m\omega+\frac{b}{2}\sx),\label{eq: H_DC}\\
    \begin{split}
            \hat H_{\rm{AC}} &= \frac{\Omega}{4} \sum_n \bigg(\cos\nu \ket{n-N_1}\bra{n}\\
            &\qquad+ \sin\nu \ket{n-N_2}\bra{n}\bigg)\otimes \sx + \rm{h.c.}\label{eq: H_AC},
    \end{split}
\end{align}
where $\hat 1_{\mathcal T}$ is the identity operator in the ${\mathcal T}$ space.

\subsection{Noise Model and Dephasing Rate}
\label{sec:lifetime}
As indicated in Fig.~\ref{fig:sketch}, the two-level system is weakly coupled to a bosonic thermal environment at temperature $T_{\rm E}$ through the $\sx$ operator. Following the noise model of Ref.~\cite{Huang2021}, we consider two different contributions to the environment spectral density, $S(\omega)=S_{\rm f}(\omega) + S_{\rm d}(\omega)$. The $1/f$ noise is given by $S_f(\omega)=|V_{\rm f}|^2 2\pi/|\omega|$. Additionally, the thermal noise, $S_{\rm d} (\omega) = \left(1+ n(\omega,T_{\rm E})\right)V_{\rm d}(\omega/2\pi)^2$, where $n(\omega,T_{\rm E}) = [e^{\omega/k_{\rm B}T_{\rm E}}-1]^{-1}$ follows the Bose-Einstein distribution.

Under these considerations, the decoherence rate approximates the following form \cite{Huang2021}
\begin{equation}
\gamma_\phi \approx 2 |V_{\rm f} |\sqrt{|\ln \omega_{\rm ir}\tau|} |g_{0\phi}|
+ \sum_{k\neq 0} 2|g_{k\phi}|^2S(k\omega),
\label{eq:gamphi}
\end{equation} 
where $\omega_{\rm ir}$ is an infrared cutoff frequency and $\tau$ is a finite characteristic time, introduced to regularize the divergence of $1/f$ noise spectrum at $\omega=0$~\cite{kou2017}. For numerical simulations~\cite{Huang2021,kou2017}, we set $V_{\rm f} = 9.0\times10^{-6}w_{\rm q}$, $V_{\rm d} = 3\times 10^{-6}w_{\rm q}$ and $\sqrt{|\ln\omega_{\rm ir}\tau|}=4$. To emphasize thermal noise effects, most significant when the drive frequency is comparable to $k_{\rm B} T$, we set $ \qubitw / k_{\rm B} T = 1.43 $, typical for low-frequency qubits \cite{Martinis_decoherence_superconducting_circ, noise_long, fast_flux}.

The properties of the Floquet qubit (and subsequently the impact of the bichromaticity) enter Eq.~(\ref{eq:gamphi}) through the weights $g_{k\phi}$, which for $\sigmax$ type system-environment interaction are defined through
\begin{align}
g_{k\phi}=\frac{1}{2T}\int_0^T e^{-ik\omega t}{\rm Tr}_{\rm S}\left[\sigmax \hat c_{\phi}(t)\right]dt,\label{eq:gkphi}
\end{align}
where $\hat c_\phi = \ket{u_+}\bra{u_+}-\ket{u_-}\bra{u_-}$. The first term on the right-hand side of Eq. (\ref{eq:gamphi}) dominates the decoherence rate in the low temperature regime. Thus, the decoherence lifetime of a Floquet qubit is largely determined by $g_{0\phi}$, which obeys the relation \cite{Huang2021}
\begin{align}
    \frac{\partial \equasi}{\partial b} = g_{0\phi},
\end{align}
where $\Delta\epsilon = \epsilon_+ -\epsilon_-$ is the Floquet quasienergy gap. The dominating term of Eq.~(\ref{eq:gamphi}) can then be re-expressed as $|\partial_b\Delta \epsilon|V_f\sqrt{|\ln\omega_{\rm{ir}}\tau|}$ and may be understood as quasienergy shifts due to uncontrolled fluctuations in the parameter $b$ (low frequency noise). However, when $\partial_b \equasi =0$, the dephasing rate will be entirely determined by the quasienergy sensitivity to broadband noise ($S(k\omega)$, $k\neq 0$). The sensitivity to AC amplitude noise $(\partial_\Omega \equasi)$ can be particularly significant experimentally, and is related to the instrumentation noise floor and control line attenuation \cite{fried2019assessing}. AC amplitude noise enters Eq~(\ref{eq:gamphi}) through the terms $S(N_1\omega)$, $S(N_2\omega)$, and the sensitivity $\partial_\Omega \equasi$ through their corresponding weights $g_{N_1\phi}, g_{N_2\phi}$. In the following, we will study the effect of both DC and AC noise on the decoherence lifetime of a solid-state qubit.

\section{Dynamical sweet and sour manifolds}

 \begin{figure}[t]
    \centering
    \includegraphics[width=0.9\linewidth]{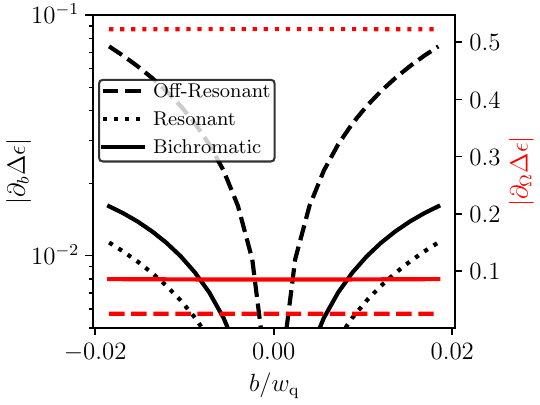}
    \caption{On the left axis (black curves), the DC sensitivity of the quasienergy gap, $\partial_b\Delta \epsilon$, is plotted as a function of the DC bias $b$ away from the static sweet spot. On the right (red horizontal lines), the sensitivity to drive amplitude noise $\partial_\Omega \equasi$ is plotted as a function of the same. Curves calculated for quasienergies generated by a drive with parameters $w=w_{\rm q}$, $N_1=3$, $N_2=1$, $\Omega=0.1\qubitw$, and different mixing angles for each curve. The dashed curve corresponds to a fast drive, set by $\nu=0$. Dotted curve corresponds to a resonant drive, set by $\nu=\pi/2$. The solid curve corresponds to a bichromatic drive set by $\nu=\pi/30$.}
    \label{fig:g0phi(b)}
\end{figure}

\label{sec:sweet_and_sour}
The regions where the decoherence lifetime $(T_\phi = \gamma_\phi^{-1})$ is high are known in the literature as {\em dynamical sweet spots} \cite{Huang2021,valery2022dyn,rigetti_2019,rigetti_bi,floquet_exp_onetone}. Given that the dominating term in 
Eq.~(\ref{eq:gamphi}) is proportional to $|\partial_b \equasi|$ (the DC noise sensitivity), $T_\phi$ will achieve its maximal values along the level curves in the drive parameters where $\partial_b\equasi\approx 0$, as discussed by Ref.~\cite{Huang2021}. However, as discussed in the previous section, $|\partial_b \equasi|$ does not uniquely determine the dephasing rate, making $\partial_b \equasi\to 0$ a necessary, but not sufficient condition for an optimal working point. Under our noise model, we find that $T_\phi$ at or near the dynamical sweet spots is inherently limited by the ``width" of the dynamical sweet region in the parameter $b$. In other words, the quality of a dynamical sweet spot depends on how broad the region is where $\partial_b \equasi \approx 0$. Moreover, in these regions of low sensitivity to DC noise, high frequency noise becomes relevant ($S(k\omega),k\neq 0$) to the decoherence lifetime. The dominant contributions to the decoherence rate will be at the fundamental drive frequencies $N_1\omega$ and $N_2\omega$~\cite{fried2019assessing}, which will affect $T_\phi$ with a distinct sensitivity to drive amplitude noise ($\partial_\Omega \Delta \epsilon$). This sensitivity should be determined by the weights $g_{N_1\phi}$ and $g_{N_2\phi}$ according to Eq.~(\ref{eq:gamphi}). We name regions with low DC noise sensitivity but high AC noise sensitivity as {\em dynamical sour spots}.

\subsection{Weak Driving Regime}
\label{sec:weak}
\begin{figure*}
    \centering
    \includegraphics[width=\linewidth]{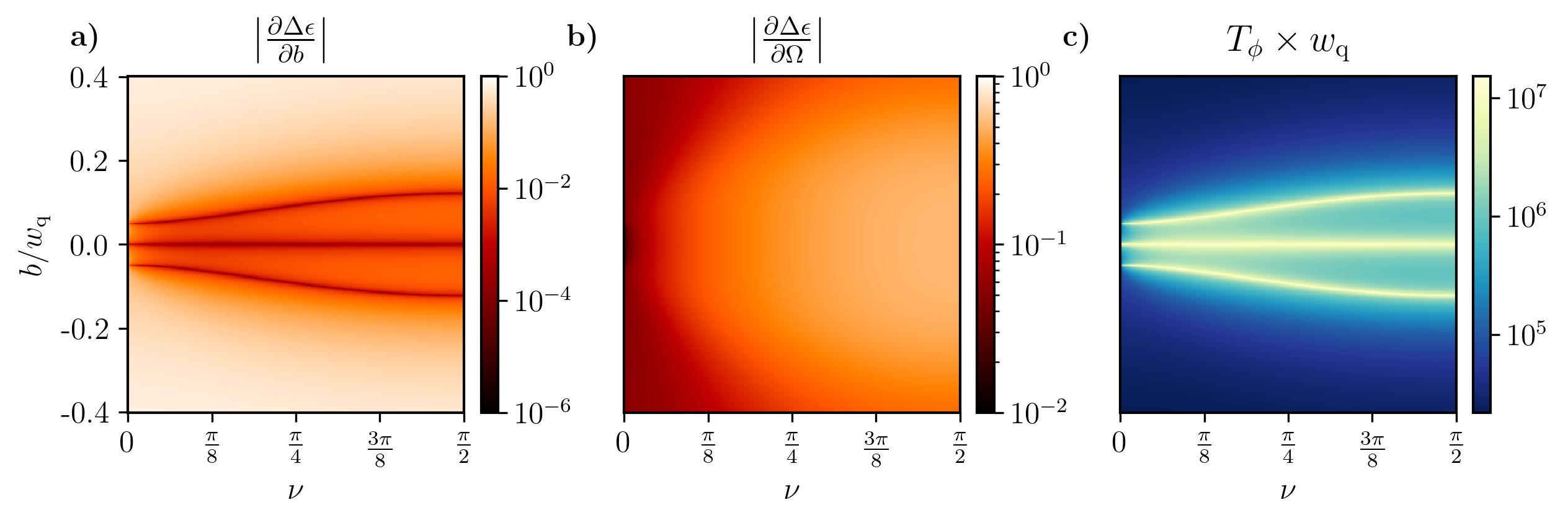}
    \caption{Quasienergy gap sensitivities to (a) DC noise ($\partial_b\Delta \epsilon$), (b) AC noise ($\partial_\Omega\Delta \epsilon$), and (c) resulting decoherence lifetime $(T_\phi)$ plotted as functions of the mixing angle $\nu$ and the DC drive strength $b$, and fixed parameters $\Omega=0.1w_{\rm q}$, $\omega = w_{\rm q}$, $N_1=3$, $N_2=1$. The bichromatic case $\nu = \pi/30$ used in Fig.~\ref{fig:g0phi(b)} lies within a narrow region of low AC sensitivity near $\nu\approx 0$.}
    \label{fig: Sw_Swr_heatmaps}
\end{figure*}

 As shown in Ref.~\cite{Huang2021}, in the weak drive regime and for $b\approx 0$, we expect dynamical sweet spots near avoided crossings which occur around $\omega \approx kw_{\rm q}$. For a monochromatic Floquet qubit, setting $k=1$ corresponds to near-resonant driving, while $k>1$ corresponds to off-resonant driving. Hence, for numerical simulations, we choose a bichromatic drive (Eq.~\ref{eq: drive def})
\begin{equation}
d(t) = \Omega\left[\cos\nu\cos(3 \qubitw t) + \sin\nu\cos(\qubitw t)\right] + b,
\label{eq:drive_model}
\end{equation}
with $\omega = \qubitw$, $N_1=3$, $N_2=1$, $\nu \in [0,\pi/2]$ and $b/w_{\rm q}\in [-1,1]$. In Eq.~(\ref{eq:drive_model}), $\nu=0$ corresponds to purely off-resonant driving ($N_1=3$), and $\nu = \pi/2$ corresponds to near-resonant driving ($N_2=1$). In the following, we will show that the properties of dynamical sweet spots generated by resonantly driven qubits (related to spin locking; see Refs.~\cite{Huang2021,noise_ithier,Guo_dephasing,Oliver_2013_noise_spectrum_analyser,decoherence_electrical_driving_2014,Smirnov_2003}) are markedly different from those produced by fast, off-resonant drives. Further, we will study trade-offs by leveraging the distinct properties of both types of drives.

In Fig~\ref{fig:g0phi(b)}, we plot $\partial_b{\Delta \epsilon}$ (on the left axis) as a function of the DC bias ($b$) on a logarithmic scale. The horizontal lines give the sensitivity to the AC noise, $\partial_\Omega \Delta \epsilon$ (on the right axis). In the off-resonant case ($\nu=0$), the sensitivity to the change in DC bias is the largest, as evidenced by higher values of $\partial_b\Delta \epsilon$ away from the narrow minimum corresponding to a dynamical sweet spot (black dashed curve). In contrast, the monochromatic resonant case ($\nu=\pi/2$) shows reduced DC sensitivity and a wider minimum (black dotted curve). However, the sensitivities are reversed for AC noise: the resonant case suffers from a large AC noise sensitivity (red dotted line) corresponding to a dynamical sour spot, whereas the off-resonant case has a lower AC noise sensitivity (red dashed line). For monochromatic drives, this sensitivity trade-off is unavoidable. 


Bichromatic drives can avoid this trade-off by balancing the two sensitivities. We find that the bichromatic drive exhibits lower DC sensitivity (black solid curve) while maintaining a wider minimum and reducing AC sensitivity (red solid curve) relative to the resonant regime. As a result, a bichromatically driven Floquet qubit demonstrates improved robustness against quasienergy gap fluctuations compared to both resonant and off-resonant monochromatic drives.

The dynamical sweet and sour spots are further explored in Fig.~\ref{fig: Sw_Swr_heatmaps}, which shows (a) the DC sensitivity ($\partial_b{\Delta \epsilon}$), (b) the AC sensitivity ($\partial_\Omega{\Delta \epsilon}$), and (c) the decoherence lifetime ($T_\phi$) as functions of the DC bias ($b$) and the mixing angle ($\nu$). Fig.~\ref{fig: Sw_Swr_heatmaps}(a) reveals three distinct dynamical sweet spot manifolds across all mixing angles. Fig.~\ref{fig: Sw_Swr_heatmaps}(b) highlights a large dynamical sour region that spans both resonant and bichromatic driving regimes, as indicated by the orange area corresponding to $\partial_\Omega \Delta \epsilon \gtrsim 10^{-1}$. The decoherence lifetime in Fig~\ref{fig: Sw_Swr_heatmaps}(c) has qualitative behavior closely following the DC noise sensitivity. Notably, along the dynamical sweet manifolds where $\partial_b \Delta \epsilon \rightarrow 0$, AC noise sensitivity becomes the dominant source of decoherence and sets the limit for the optimal decoherence lifetime. As a result, $T_\phi$ reaches an upper bound of approximately $1.5\times10^7w_{\rm q}^{-1}$. 

We also emphasize that the numerically implemented noise model described following Eq.~(\ref{eq:gamphi}) accounts only for $1/f$ and thermal noise. In practice, additional noise sources, particularly instrumentation noise, have been shown to significantly affect the decoherence lifetime of parametrically driven qubits~\cite{fried2019assessing}. Instrumentation noise will significantly increase the deleterious effect of dynamical sour spots shown in Fig.~\ref{fig: Sw_Swr_heatmaps}(b). This added impact of instrumentation noise falls outside the scope of this work and will be addressed in future research.

\subsection{$T_\phi$ Optimization Beyond Weak Driving Regime}
\label{sec:beyond_weak}


\begin{figure*}
    \centering
    \includegraphics[width=\linewidth]{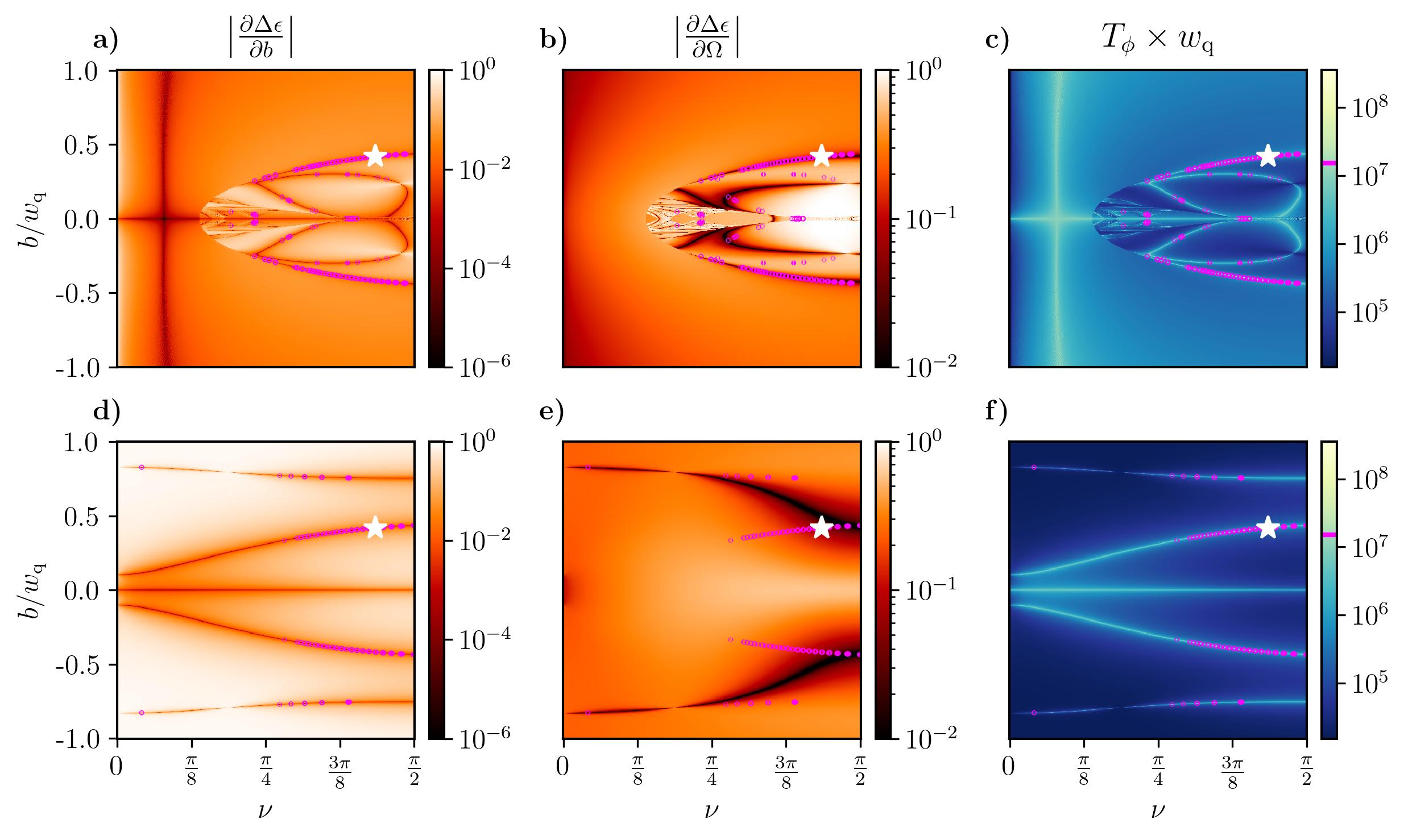}
    \caption{Quasienergy gap sensitivities to (a,d) DC noise ($\partial_b\Delta \epsilon$), (b,e) AC noise ($\partial_\Omega\Delta \epsilon$), and (c,f) resulting decoherence lifetime $(T_\phi)$ plotted as functions of the mixing angle $\nu$ and the DC drive strength $b$, and fixed parameters $\Omega=0.1w_{\rm q}$, $N_1=3$, $N_2=1$. For (a-c), the base drive frequency $\omega$ is optimized for each $(b,\nu)$ to $ \omega^* = \Theta$, defined below Eq.~(\ref{eq:gap_mmf}). For (d-f), the base drive frequency is fixed to a single optimal $\omega^*=0.89 w_{\rm q}$ corresponding to the white stars in (a-c), chosen in a doubly sweet region with low DC and AC noise sensitivities. 
    }
    \label{fig: Sw_Swr doubly_sweet_heatmaps}
\end{figure*}

In the weak driving regime assuming $\omega= w_{\rm q}$, we identified regions of low DC and AC sensitivity; however, the overlap between these regions was limited due to the enhanced AC noise sensitivity induced by the resonant drive. In this subsection, we move beyond the weak driving regime, taking into account the observation in Ref.~\cite{Huang2021} that for intermediate drive strengths, the base driving frequency should be detuned from resonance to maximize $T_\phi$. As such, we identify optimal base drive frequencies $(\omega^*)$ that yield larger regions of overlap between low DC and AC noise sensitivities. 

We can derive an approximate expression for the quasienergy gap assuming neither drive is slow compared to the qubit gap ($\omega \geq \omega_{\rm q}$) and that one drive is significantly weaker than the other, i.e., with a small mixing angle ($\nu\ll \pi/4$) as in Fig.~(\ref{fig:g0phi(b)}) (see App.~\ref{sec:quasienergy} for details)
\begin{align}
    \Delta \epsilon &=  \sqrt{\left(\omega - \Theta\right)^2+\left( w_{\rm q} \tilde J_{1,1}\tilde J_{1,2}\right)^2 }.
    \label{eq:gap_mmf}
\end{align}
We define $\Theta^2 = b^2 + \left( w_{\rm q} \tilde J_{0,1}\tilde J_{0,2}\right)^2  $, $\tilde{J}_{l,1} = J_l\left(\frac{\Omega \cos \nu}{N_1 \omega}\right)$ and $\tilde{J}_{l,2} = J_l\left(\frac{\Omega \sin \nu}{N_2 \omega}\right)$, where $J_l$ is $l^{\rm th}$ order Bessel function of first-kind. Although the above expression deviates from the exact quasienergy gap beyond the small mixing angle assumption, it still accurately predicts the positions of the minima and maxima of the quasienergy gap. We exploit this property of Eq.~(\ref{eq:gap_mmf}) to identify optimal frequencies that minimize $\partial_b\equasi$. The differential of the gap with respect to the DC bias strength is given by
\begin{equation}
    \frac{\partial \Delta \epsilon}{\partial b} = \frac{b({\omega}/{\Theta} - 1)}{\Delta \epsilon}.
    \label{eq:bias_sensivity}
\end{equation}
Eq.~(\ref{eq:bias_sensivity}) sets $\partial_b \Delta \epsilon = g_{0\phi}$ as directly proportional to $b$. Further, the sensitivity to the bias vanishes when $\omega = \Theta$, which in the weak driving and small DC bias regime corresponds to $\Theta \approx w_{\rm q}$. This requirement is satisfied by including a resonant component in the drive, which agrees with our previous analysis. Note that the DC noise sensitivity diverges whenever the quasienergy gap goes to 0.

In Fig.~\ref{fig: Sw_Swr doubly_sweet_heatmaps} (a-c), we set the drive frequency to $\omega \equiv \omega^* (b,\nu) = \Theta$ keeping $\Omega = 0.4 w_{\rm q}$, $N_1=3$ and $N_2=1$ fixed. We analyze the DC (Fig.~\ref{fig: Sw_Swr doubly_sweet_heatmaps} (a)) and AC (Fig.~\ref{fig: Sw_Swr doubly_sweet_heatmaps} (b)) noise sensitivities of the quasienergy gap as functions of DC bias strength ($b$) and mixing angle ($\nu$). Compared to Fig.~\ref{fig: Sw_Swr_heatmaps}, we obtain a significantly richer landscape of noise-insensitive regions, highlighted by the darker areas in panels (a) and (b). AC noise sensitivity ($\partial_\Omega \equasi$) still remains pronounced in the near-resonant drive dominated regime (e.g., the whiter region near $\nu\approx \pi/2$ and $b\approx 0$ in Fig.~\ref{fig: Sw_Swr doubly_sweet_heatmaps}(b)). However, unlike Fig.~\ref{fig: Sw_Swr_heatmaps}(b) with a non-optimal $\omega$, we find regions of low AC noise sensitivity even for regimes dominated by the near-resonant drive. Moreover, we identify bichromatic regions where both DC and AC noise sensitivities are simultaneously small. These {\em doubly sweet spots}, marked by pink dots in Fig.~\ref{fig: Sw_Swr doubly_sweet_heatmaps}(c), correspond to enhanced decoherence lifetime $T_\phi$ exceeding the maximal $T_\phi$ obtained in Fig.~\ref{fig: Sw_Swr_heatmaps}(c) $(1.5\times10^7w_{\rm q}^{-1})$. Fig.~\ref{fig: Sw_Swr doubly_sweet_heatmaps}(a) also shows an additional sweet spot manifold $\nu\approx \pi/12$, for a wide range of bias values ($b$), but this manifold remains sensitive to AC noise. 


Experimentally, it can be difficult to change the base drive frequency $\omega$ as needed for obtaining optimal $T_\phi$ in Fig.~4(c). As such, we also fix a single optimal base frequency ($\omega^*$) on one of the doubly sweet spot manifolds (white star in Fig.~\ref{fig: Sw_Swr doubly_sweet_heatmaps} (a-c)) and vary $b$ and $\nu$ to produce Figs.~\ref{fig: Sw_Swr doubly_sweet_heatmaps}(d-f). Compared to Figs.~\ref{fig: Sw_Swr_heatmaps}(a-c),  
Figs.~\ref{fig: Sw_Swr doubly_sweet_heatmaps}(d-e) show additional DC and AC sweet spot manifolds, including a doubly sweet spot manifold containing the chosen optimal base drive frequency. Even with fixed $\omega$, tunability along the doubly sweet spot manifold is preserved. We also note that Fig.~\ref{fig: Sw_Swr doubly_sweet_heatmaps}(e) shows a broader region of AC insensitivity  than either Fig.~\ref{fig: Sw_Swr doubly_sweet_heatmaps}(b) or Fig.~\ref{fig: Sw_Swr_heatmaps}(b). 

\subsection{Fast Driving Regime: GVV Perturbation Theory}
\label{sec:res_singl}
In the preceding subsections, we examined the decoherence lifetime under weak and beyond-weak driving conditions, focusing on near-resonant regimes. In this subsection, we extend our analysis to the fast-driving regime ($\omega \gg w_{\rm q}$), where we explore both the quasienergy gap and the resulting decoherence lifetime. In this regime, neither drive is resonant with the qubit gap $w_{\rm q}$ and the strong sensitivity to AC noise observed in previous sections is negligible. Therefore, we will constrain our analysis to the DC noise sensitivity $\partial_b \equasi$. Building on our earlier results, we leverage the structure of the quasienergy gap to identify parameter regimes with high decoherence lifetime. In particular, we analyze scenarios where the Floquet states become nearly degenerate, specifically when $b \approx k \omega$~\cite{son2009floquet}, where $k=mN_1 + lN_2$. We define $\delta$ through $b = k\omega + \delta$, with $0< |\delta|\ll 1$. The RWA quasienergy gap takes the following form 
\begin{equation}
{\Delta\epsilon_{\rm RWA} = \sqrt{\delta^2 +\left( w_{\rm q} \tilde J_{-m,1}\tilde J_{-l,2}\right)^2 }},
\end{equation}
obtained from the effective $2 \times 2$ Hamiltonian
\begin{equation}
    H_{\rm RWA}=\begin{pmatrix}
            -\frac{b}{2} & -\frac{w_{\rm q}}{2} \tilde{J}_{-m,1}\tilde{J}_{-l,2}\\[4pt]
            -\frac{w_{\rm q}}{2}\tilde{J}_{-m,1}\tilde{J}_{-l,2} & \frac{b}{2} -mN_1\omega - l N_2 \omega
            \end{pmatrix},
\label{eq:HRWA}            
\end{equation}
obtained in the rotating wave approximation (RWA), with $\tilde{J}_{l,1}$ and $\tilde{J}_{l,2} $ as defined below Eq.~(\ref{eq:gap_mmf}).

To calculate the AC Stark shift correction $(\chi)$ to the RWA quasienergy gap, we apply the GVV nearly-degenerate perturbation theory~\cite{son2009floquet}, accounting for the influence of levels beyond those selected by the degeneracy condition. We employ the GVV method considering higher-order perturbations in $w_{\rm q}/\omega$ in contrast to RWA, where only the zeroth-order perturbative effect is considered. The effective $2\times 2$ GVV Hamiltonian includes the AC Stark shift $\chi$
\begin{equation}
         H_{\rm GVV}=
            \begin{pmatrix}
             -\frac{b}{2} +\chi & -\frac{w_{\rm q}}{2} \tilde{J}_{-m,1}\tilde{J}_{-l,2}\\[4pt]
            -\frac{w_{\rm q}}{2} \tilde{J}_{-m,1}\tilde{J}_{-l,2} & \frac{b}{2} -mN_1\omega - lN_2 \omega -\chi
            \end{pmatrix},
\label{eq:HGVV}            
\end{equation}
with the quasienergy gap given by the relation
\begin{equation}
\Delta \epsilon_{\rm GVV} = \sqrt{\Delta\epsilon_{\rm RWA}^2 + 4\chi\delta + 4\chi^2}.\label{eq:equasi GVV}
\end{equation}

\begin{figure}[t]
\includegraphics[width=\columnwidth]{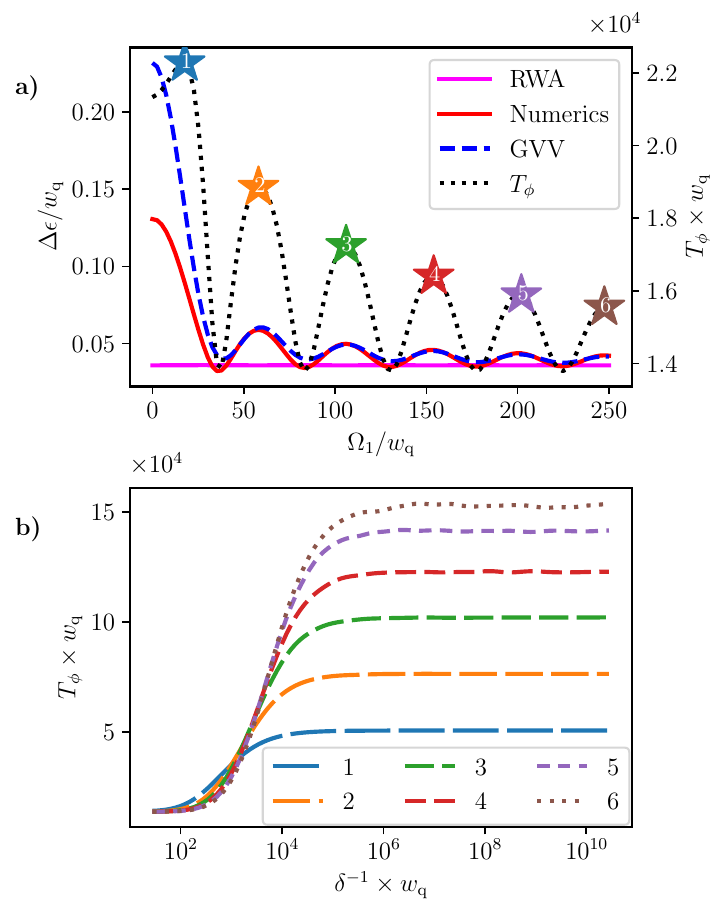}
 \caption{(a) On left axis, Floquet quasienergy gap plotted as a function $\Omega_1 = \Omega \cos(\nu)$, where $\Omega_2 = \Omega\sin(\nu)$ is held constant.  We compare the exact numerical results (solid red line) with the RWA (solid purple) and GVV (dashed blue) calculations. On the right axis, the decoherence lifetime for the same system. Numbered stars correspond to $T_\phi$ local maxima. (b) $T_\phi$ plotted as a function of $\delta$ at each local maxima indicated in (a). Parameters: $N_1 = 3$, $N_2= 1$, $m=1$, $l=-2$, $\Omega_2=\qubitw$ and $\omega = 10 w_{\rm q}$. The bias $b$ is set by the nearly degenerate condition and $\delta$. In (a) $\delta=0.01\qubitw$.}
\label{fig:rwa_gvv_Tphi}
\end{figure}

 Using the GVV method, we obtain the following expressions for the AC Stark shift up to the leading order (see App.~\ref{sec:lev_shift}),
\begin{align}
    \chi&=\sum_{\substack{j,p=-\infty\\j,p\neq -m,-l}}^{\infty} \frac{-\left(\tilde{J}_{j,1}\tilde{J}_{p,2}\right)^2}{4(b +jN_1\omega+pN_2\omega)}{w_{\rm q}^2} + \mathcal{O}\left(\frac{w_{\rm q}}{\omega}\right)^3.
\label{eq:Stark_Shift_onequbit}
\end{align}

Using Eq.~(\ref{eq:equasi GVV}), the DC bias sensitivity is given by 
\begin{align}
    \frac{\partial \equasi_{\rm{GVV}}}{\partial b} &= \frac{1}{\equasi_{\rm{GVV}}}(\delta + 2\chi)\left(1+2\frac{\partial \chi}{\partial b}\right).\label{eq:equasi GVV b sensitivity}
\end{align}
We find that in the fast driving regime ($\omega/w_{\rm q}\gg 1$), $\partial_b \chi \ll 1$. Hence, Eq.~(\ref{eq:equasi GVV b sensitivity}) reduces to
\begin{align}
    \frac{\partial \equasi_{\rm{GVV}}}{\partial b} &\approx \frac{1}{\equasi_{\rm{GVV}}}(\delta +  2\chi).\label{eq:equasi GVV b sensitivity2}
\end{align}
Figure~\ref{fig:rwa_gvv_Tphi} (a) shows the dependence of the quasienergy gap on one of the drive amplitudes $\Omega_1 \equiv \Omega\cos\nu$, keeping the other drive amplitude fixed and weak compared to $\omega$, $\Omega_2 \equiv \Omega\sin\nu = 0.1\omega$. We compare the quasienergy gaps obtained from the RWA (solid horizontal purple line) and the GVV (dashed blue curve) Hamiltonians to the exact values calculated numerically (solid red curve). The GVV result, which incorporates the AC Stark shift $\chi$, shows markedly better agreement with numerical simulations than RWA, with disagreement when $\Omega_1 \leq \omega$. 
For fixed $\delta>0$, the DC noise sensitivity is primarily determined by the magnitude of the quasienergy gap and the AC Stark shift $\chi$ (see Eq.~(\ref{eq:equasi GVV b sensitivity2})); a larger quasienergy gap or AC Stark shift (since $\chi<0$) leads to lower noise sensitivity.  The AC Stark shift follows the minima and maxima of the quasienergy gap. Consequently, the decoherence lifetime (black dotted curve) reaches its minima and maxima (indicated by stars) at the corresponding minima and maxima of the quasienergy gap, respectively.

To further analyze these maxima, in Fig.~\ref{fig:rwa_gvv_Tphi}(b), we plot the decoherence lifetimes, using colors corresponding to each star in Fig.~\ref{fig:rwa_gvv_Tphi}(a), as a function of $\delta^{-1}$. We notice from Eq.~(\ref{eq:equasi GVV b sensitivity2}) that the DC noise sensitivity can be minimised by making $\delta$ comparable to $2\chi$. For large $\delta$ (small $\delta^{-1}$), the aforementioned compensation results in a monotonous increase of decoherence lifetime as a function of $\delta^{-1}$. However, as $\delta\rightarrow 0$ (large $\delta^{-1}$), the decoherence lifetime saturates as both the quasienergy gap and the AC Stark shift become independent of $\delta$ and take a constant value.

\section{Conclusion and Outlook}
\label{sec:conclusion}
Our work establishes bichromatic driving as a powerful strategy to engineer dynamical sweet manifolds in solid-state driven qubits, achieving high decoherence lifetimes while maintaining tunability. By combining near-resonant and off-resonant drives, we suppress sensitivity to low-frequency (e.g., $1/f$) noise by redirecting the dominant sources of decoherence to frequency regions where environmental noise is minimal. Crucially, we identify not only high-coherence ``sweet spots" but also  ``sour spots" where the quasienergy gap is still sensitive to noise at the drive frequency. These sour spots are prominent when one of the drives is nearly resonant with the DC qubit gap, underscoring the need for careful frequency selection in multi-tone protocols beyond minimizing the quasienergy sensitivity to DC bias noise ($\partial_b \Delta \epsilon$). Hence, the interplay of sweet and sour spot dynamics provides a framework to balance tunability and decoherence lifetime, which is critical for single and multi-qubit operations.

Central to our work is the derivation of analytic expressions for the sensitivity of the Floquet quasienergy gap to external drives in the intermediate strength driving regime and fast driving regime. We employed the derived analytic expressions to connect the drive parameters to the decoherence lifetime, explaining the emergence of dynamical sweet and sour spots. The analytic expressions were further employed to find optimal drive frequencies, giving a long decoherence lifetime.


Future work will investigate the application of bichromatic Floquet engineering for single and multi-qubit gate operations. We would also like to study the application of Floquet engineering in multi-qubit architectures, suppressing crosstalk while preserving tunable interactions. It would also be important to investigate how instrumentation noise affects the structure and effectiveness of the dynamical sweet and sour manifolds. Experimental validation of our results in circuit QED platforms will bridge theory and device-specific noise landscapes to advance decoherence mitigation in parametrically driven solid-state quantum architectures.
\section*{ACKNOWLEDGMENTS}

We thank Abhishek Chakraborty, Nicolas Didier, and Rosario Fazio for the valuable discussions. This work was supported by the U. S. Army Research Office under grant W911NF-22-1-0258. DD acknowledges support by PNRR MUR Project No. PE0000023-NQSTI (National Quantum Science and Technology Institute). Y.K. is supported by the National Research Foundation of Korea (NRF) grant funded by the Korea government (MSIT) (Nos. RS-2024-00353348 and RS-2023-NR068116).

\appendix

\section{Quantum Floquet Theory}
\label{sec:floquet}
We consider a time-dependent, periodic Hamiltonian with period $T$. The corresponding Schr\"odinger equation (setting $\hbar = 1$),
\begin{equation}
    i\frac{\partial}{\partial t} \ket{\psi} = \hat H(t) \ket \psi, \label{schrodinger eq}
\end{equation}
does not admit any stationary eigenstates. It does, however, admit orthonormal, quasi-stationary solutions called Floquet states
\begin{equation}
    \ket{\psi_\pm(t)} = e^{-i\epsilon_\pm t}\ket{u_\pm(t)},\label{floquet_state}
\end{equation}
where $\epsilon_\pm$ is called the quasi-energy, and $\ket{u_\pm(t)}$ is the $T$-periodic Floquet mode. Note that the quasi-energies are defined only up to additive factors of $\omega = 2\pi/T$, such that $\epsilon_\pm \equiv \epsilon_\pm + k\omega$ for any integer $k$. The quasienergies and their respective Floquet modes are eigenvalues and eigenvectors of a Hermitian operator called the Floquet Hamiltonian
\begin{align}
    \hat H_{\rm{F}} (t)= \hat H(t) - i\frac{\partial}{\partial t}.
\end{align}

The problem of finding the time-dependent dynamics may thus be recast as an eigenvalue problem for a time-dependent operator. We may further reduce the problem to a time-independent eigenvalue problem by exploiting the periodicity of the Floquet modes. The modes may be expanded as
\begin{equation}
    \ket{u_\pm} = \sum_{n,\alpha} c^\pm_{n,\alpha}e^{in\omega t}\ket{\alpha}, \label{eq: time-dependent floquet modes}
\end{equation}
where $c^\alpha_{n,\pm} = \frac {1} {\rm T} \int_0^T e^{-in\omega t} \langle \alpha | u_\pm\rangle dt$ are the time-independent Floquet coefficients. The set of products $e^{in\omega t}\ket \alpha$ satisfies the properties of a product space $\mathcal T \otimes \mathcal H$ between the space of T-periodic functions $\mathcal T$ and atomic states of the undriven system $\mathcal H$. Setting the basis for $\mathcal T$ as $\left \{ \ket n = e^{in\omega t} \right\}_{n\in\mathbb Z}$, we may write the Floquet modes as
\begin{align}
    \ket {u_\pm} = \sum_{n,\alpha} c^\pm_{n,\alpha} \ket{n,\alpha}. \label{eq:floquet modes}
\end{align}

For notational clarity, we will use Greek indices to represent the Floquet modes or atomic states, while the Latin indices will be used for T-periodic functions. When written in these coordinates, called the extended Hilbert space, the Floquet Hamiltonian takes the form of an infinite-dimensional time-independent operator with components $\bra{n,\alpha} \hat H_{\rm F}\ket{m,\beta}$, eigenvalues $\epsilon_\pm^k$ and eigenvectors $\ket{u_\pm^k}$. The extended Hilbert space eigenvalue-vector pairs correspond to quasienergies and modes in Eq.~(\ref{floquet_state}) shifted in energy and frequency by $k\omega$
\begin{align}
    \epsilon^k_\pm &= \epsilon_\pm - k\omega\\
    \ket{u^k_\pm} &= \sum_{n,\alpha} c_{n,\alpha}^\pm \ket{n-1, \alpha}.
\end{align}
The physically observable modes and quasi-energies correspond to the equivalence classes over every $k$ of these eigenvalues and eigenvectors. When we require a numerical result, we will pick $k=0$, but we remark that all physically measurable quantities we report are invariant to the choice of $k$.

In this section, we utilized Shirley's Floquet theory~\cite{Shirley1965} to define the Floquet quasienergies and modes, a framework well-suited for computational analysis due to its ability to directly extend the monochromatic framework to bichromatic driving without introducing significant complexity. However, for deriving the analytic expressions governing the quasienergy gap (see App.~\ref{sec:quasienergy}) and for probing the multi-photon resonance regime (see App.~\ref{sec:lev_shift}), we will adopt multi-mode Floquet theory \cite{ho1983semiclassical}, which is more suited for analytic calculations.
\begin{widetext}
\section{Floquet Hamiltonian}
\label{sec:floq_ham_app}
In this appendix, we calculate the RHS of Eq.~(3) with Hamiltonian in Eq.~(8) of the main text. We choose our basis $\ket{m,\alpha} = e^{im\omega t}\ket{\alpha}$, and define $\bra{k,\beta}$ as the linear functional which acts on $\ket{m,\alpha}$ such that
\begin{equation}
    \bra{k,\beta}\ket{m,\alpha} = \bra{\beta}\ket{\alpha}\int_0^T dt\; e^{-ik\omega t}e^{im\omega t}.
\end{equation}
The action of $-i{\partial_t}$ on $\ket{m,\alpha}$ and $\hat{H}(t)$ is, respecttively, given by
\begin{align}
    -i\frac{\partial}{\partial t} e^{im\omega t} \ket{\alpha} &= m\omega e^{im\omega t}\ket{\alpha}=m\omega\ket{m,\alpha},
    \label{eq:floq_alpha}
\end{align}
and
\begin{align}
    \hat H(t) \ket{m,\alpha} = -\frac{w_q}{2}e^{im\omega t}\sz\ket{\alpha} + \frac{d(t)}{2}e^{imwt}\sx\ket{\alpha}.
    \label{eq:floq_Hamil}
\end{align}
Considering  the drive to be combinations of complex exponentials,
\begin{align}
    d(t) =& \frac{\Omega}{2}\left[\cos\nu\left(e^{iN_1\omega t} + e^{-iN_1\omega t}\right) + \sin\nu\left(e^{iN_2\omega t} + e^{-iN_2\omega t}\right)  \right] + {b},
\end{align}
we obtain
\begin{align}
    d(t)e^{imwt} &=\frac{\Omega}{2}\left[\cos\nu\left(e^{i(m+N_1)\omega t} + e^{i(m-N_1)\omega t}\right) +
    \sin\nu\left(e^{i(m+N_2)\omega t} + e^{i(m-N_2)\omega t}\right)\right] + e^{im\omega t}b\\
    &= \frac{\Omega}{2}\left[\cos\nu\big(\ket{m+N_1} + \ket{m-N_1}\big) + \sin\nu\big(\ket{m+N_2} + \ket{m-N_2}\big)\right] + {b}\ket{m}.
\end{align}
Thus, the action of the time-dependent Hamiltonian on a basis state is given by
\begin{align}
\begin{split}
    \hat H(t)\ket{m,\alpha } &= -\frac{w_q}{2}\ket{m} \otimes \hat \sigma_z \ket{\alpha} + \frac{b}{2}\ket{m}\otimes \hat \sigma_x \ket{\alpha}\\
    &\qquad+\frac{\Omega}{4}\bigg[ \cos\nu \bigg(\ket{m+N_1} + \ket{m-N_1} \bigg)  + \sin \nu \bigg(\ket{m+N_2} + \ket{m-N_2}\bigg)\bigg]\otimes \hat \sigma_x\ket{\alpha}.
\end{split}
\end{align}
Note that the time-dependence of the Hamiltonian has been incorporated into the basis vectors, leaving all coefficients time-independent. Hence, the Floquet Hamiltonian in the $\ket{m,\alpha}$ basis can be expressed as
\begin{align}
    \begin{split}
        \hat H(t) - i\frac{\partial}{\partial t} &= \sum_m \ket{m}\bra{m} \otimes\bigg(-\frac{w_q}{2}\sz + \frac{b}{2}\sx  + m\omega\bigg) +\frac{\Omega}{4}\bigg[ \cos\nu \bigg(\ket{m+N_1}\bra{m} + \ket{m-N_1}\bra{m}\bigg) \\
        &\qquad + \sin\nu \bigg(\ket{m+N_2}\bra{m} + \ket{m-N_2}\bra{m}\bigg) \bigg]\otimes \sx.
    \end{split} \label{app: Full H_floquet}
\end{align}
The Floquet Hamiltonian in Eq.~(\ref{app: Full H_floquet}) separates into the terms $\hat H_0, \hat H_{\rm{DC}}, \hat H_{\rm{AC}}$ defined in Eqs.~(\ref{eq: H_D}), (\ref{eq: H_DC}) and (\ref{eq: H_AC}), respectively. The Hamiltonian $\hat H_{0} + \hat H_{\rm{DC}}$ is easily diagonalized with eigenvalues (
$\lambda_{\pm}^m$) and eigenvectors $(\ket{m,\pm})$, where
\begin{align}
    \lambda_\pm^m &= m\omega  \pm \frac{1}{2} \mathcal E, \label{app: Hq eigenstates}\\
    \ket{m, +} &= \cos\theta \ket{m,g} + \sin\theta \ket{m,e},\label{app: m+}\\
    \ket{m, -} &= \sin\theta\ket{m,g} - \cos\theta\ket{m,e}.\label{app: m-}
\end{align}
$g,e$ enumerate the $\sigmax$ eigenstates, $\theta = \frac{1}{2}\tan^{-1}\left(\frac{b}{w_{\rm q}}\right)$, and $\mathcal E = \sqrt{w_{\rm q}^2 + b^2}$ is the transition frequency of the undriven qubit. In this new basis, the $\hat H_{0}+\hat H_{\rm{DC}}$, $\sx$, and $\sz$ operators reduce to
\begin{align}
\hat H_0 + \hat H_{\rm{DC}} &= \sum_m \ket{m}\bra{m} \otimes\bigg[m\omega + \frac{1}{2}\mathcal E \bigg(\ket{+}\bra{+} - \ket{-}\bra{-}\bigg)\bigg],\\
    \sz &= \begin{bmatrix}
        \cos 2\theta & \sin 2\theta \\
        \sin2\theta & -\cos 2\theta
    \end{bmatrix}, {\rm ~and~}
     \sx = \begin{bmatrix}
        \sin 2\theta & -\cos 2\theta\\
        -\cos 2\theta & -\sin 2\theta
    \end{bmatrix}. \nonumber
\end{align}

\section{Quasienergy Gap}
\label{sec:quasienergy}
In this section, we will use the multi-mode Floquet theory to calculate an analytic expression for the quasienergy gap. The multi-mode Floquet theory exploits the presence of two periodic drives of commensurate frequencies to expand the Floquet modes in terms of two periodic functions as follows,
\begin{equation}
\ket{u_\sigma (t)}=\sum_{\substack{m,l\\\alpha=\pm}} c^{\sigma,\alpha}_{ml} \ket{u^{ml}_{\alpha}},
\end{equation}
where $\sigma = \{+,-\}$ and $\ket{u_{\alpha}^{ml}} = e^{imN_1\omega t}e^{ilN_2\omega t}\ket{\alpha}$. Unlike in Shirley's formulation of Floquet theory, in multi-mode Floquet theory we treat the product $e^{imN_1\omega t}e^{ilN_2\omega t}$ as a vector in a product space $\mathcal T_1 \otimes \mathcal T_2$ of periodic functions with period $2\pi/N_1\omega$ and $2\pi/N_2\omega$, respectively. This defines 
\begin{align}
    \ket{m,l} =  e^{imN_1\omega t}e^{ilN_2\omega t} = \ket{m}\otimes \ket{l},
\end{align}
and thus $\ket{u^{ml}_\alpha} = \ket{m}\otimes \ket{l}\otimes\ket{\alpha}$ is a triple Kronecker product.
Following Ref.~\cite{son2009floquet}, we rotate the Hamiltonian $\hat H(t) = -w_{\rm q}/2\sz + d(t)/2\sx$ by a $\pi/2$ rotation around the y-axis, such that
\begin{equation}
H_{\rm q} = -\frac{w_{\rm q}}{2} \sigmax - \frac{d(t)}{2}\sigmaz.
\end{equation}
Using the multimode Floquet theory, the eigen-equation for the Floquet Hamiltonian can be written as
\begin{multline}
\left(-\frac{w_{\rm q}}{2} \sigmax-\frac{b}{2} \sigmaz+mN_1\omega+lN_2\omega\right)\ket{u_{\sigma}^{ml}}-\frac{\Omega}{4}\cos\nu\sigmaz \left(\ket{u_{\sigma}^{(m-1)l}}+\ket{u_{\sigma}^{(m+1)l}}\right)\\
-\frac{\Omega}{4}\sin \nu\sigmaz\left(\ket{u_{\sigma}^{m(l-1)}}+\ket{u_{\sigma}^{m(l+1)}}\right)
=\epsilon_\sigma\ket{u_{\sigma}^{ml}}.
\end{multline}
Hence, the Floquet Hamiltonian matrix in the basis $\mathcal T_1 \otimes \mathcal T_2$ would be given by
\setcounter{MaxMatrixCols}{20}
\begin{equation}
\hat H_{\rm F}=
\begin{bmatrix}
\ddots & ~ & ~ & ~ & ~ & ~ & ~ & ~ & ~ & ~ & ~ & ~ & ~ \\
~ & M_{-1,-1} & R & 0 & \cdots & D & 0 & 0 & \cdots & 0 & 0 & 0 & ~\\
~ & R & M_{-1,0} & R & \cdots & 0 & D & 0 & \cdots & 0 & 0 & 0 & ~\\
~ & 0 & R & M_{-1,+1} & \cdots & 0 & 0 & D  & \cdots & 0 & 0 & 0 & ~ \\
~ & \vdots & \vdots & \vdots & \ddots & \vdots & \vdots & \vdots & \ddots & \vdots & \vdots & \vdots & ~ \\
~ & D & 0 & 0 & \cdots & M_{0,-1} & R & 0 & \cdots & D & 0 & 0 & ~\\
~ & 0 & D & 0  & \cdots & R & M_{0,0}&R  & \cdots & 0 & D  & 0 & ~\\
~ & 0 & 0 & D   & \cdots & 0 & R & M_{0,+1}  & \cdots & 0 & 0 & D  & ~ \\
~ & \vdots & \vdots & \vdots & \ddots & \vdots & \vdots & \vdots & \ddots & \vdots & \vdots & \vdots & ~ \\\
~ & 0 & 0 & 0 & \cdots &D & 0 & 0 & \cdots & M_{+1,-1} & R & 0 & ~ \\
~ & 0 & 0 & 0 & \cdots & 0 & D & 0  & \cdots & R & M_{+1,0}&R  & ~\\
~ & 0 & 0 & 0 & \cdots & 0 & 0 & D   & \cdots & 0 & R & M_{+1,+1}  & ~ \\
~ & ~ & ~ & ~ & ~ & ~ & ~ & ~ & ~ & ~ & ~ & ~ & \ddots \\
\end{bmatrix},
\label{eq:bigmat}
\end{equation}
where
\begin{equation}
\label{eq:bigmat_M}
M_{ml}=
\begin{bmatrix}
-\frac{b}{2}+mN_1\omega+lN_2\omega & -\frac{w_{\rm q}}{2}\\
-\frac{w_{\rm q}}{2} & \frac{b}{2}+mN_1\omega + lN_2\omega
\end{bmatrix},
\end{equation}
and
\begin{equation}
\label{eq:bigmat_DR}
D=\begin{bmatrix}
    -\frac{\Omega}{4} \cos \nu & 0\\
    0 & \frac{\Omega}{4} \cos \nu
\end{bmatrix};
R=\begin{bmatrix}
    -\frac{\Omega}{4} \sin \nu & 0\\
    0 & \frac{\Omega}{4} \sin \nu
\end{bmatrix}.
\end{equation}
Taking the energy gap $-w_{\rm q}$ as the perturbation parameter, we divide the Floquet matrix into two parts: an unperturbed part $H_{0,{\rm F}}$ and a perturbed part $(-w_{\rm q}/2) V^\prime$, 
\begin{equation}
    H_{\rm F}=H_{0,{\rm F}}+\frac{(-w_{\rm q})}{2}V^\prime.
\end{equation}
The unperturbed part of the Floquet matrix can thus be written as
\begin{multline}
    H_{0,{\rm F}}=\\
    \scriptsize{
    \begin{bmatrix}
\ddots & ~ & ~ & ~ & ~ & ~ & ~ & ~ & ~ & ~ & ~ & ~ & ~ \\
~ & -\frac{b}{2}-N_\Sigma\omega & 0 & -\frac{\Omega}{4}\sin \nu & 0 & 0 & 0 & \cdots & -\frac{\Omega}{4} \cos \nu & 0 & 0 & 0 & ~\\
~ & 0 & \frac{b}{2}-N_\Sigma\omega & 0 & \frac{\Omega}{4} \sin \nu & 0 & 0 & \cdots & 0 & \frac{\Omega}{4} \cos \nu& 0 & 0 & ~\\
~ & -\frac{\Omega}{4} \sin \nu & 0 & -\frac{b}{2} -N_1\omega & 0 & -\frac{\Omega}{4} \sin \nu & 0 & \cdots  & 0 & 0 & -\frac{\Omega}{4} \cos \nu & 0 & ~ \\
~ & 0 & \frac{\Omega}{4} \sin \nu & 0 & \frac{b}{2}-N_1\omega & 0 & \frac{\Omega}{4} \sin \nu & \cdots & 0 & 0 & 0 & \frac{\Omega}{4} \cos \nu & ~ \\
~ & 0 & 0 & -\frac{\Omega}{4} \sin \nu & 0 & -\frac{b}{2}-N_\Delta\omega & 0 & \cdots & 0 & 0 & 0 & 0 & ~\\
~ & 0 & 0 & 0  & \frac{\Omega}{4} \sin \nu & 0 & \frac{b}{2}-N_\Delta\omega &\cdots  & 0 & 0 & 0  & 0 & ~\\
~ & \vdots & \vdots & \vdots & \ddots & \vdots & \vdots & \vdots & \ddots & \vdots & \vdots & \vdots & ~ \\\
~ & -\frac{\Omega}{4}\cos\nu & 0 & 0 & 0 & 0 & 0  & \cdots & -\frac{b}{2}-N_2\omega  & 0 & -\frac{\Omega}{4}\sin\nu & 0  & ~ \\
~ & 0 & \frac{\Omega}{4}\cos\nu & 0 & 0 & 0 & 0 & \cdots & 0 & \frac{b}{2}-N_2\omega & 0 & \frac{\Omega}{4}\sin\nu & ~ \\
~ & 0 & 0 & -\frac{\Omega}{4}\cos\nu & 0 & 0 & 0 & \cdots  & -\frac{\Omega}{4}\sin\nu & 0 & -\frac{b}{2} & 0  & ~\\
~ & 0 & 0 & 0 & \frac{\Omega}{4}\cos\nu & 0 & 0 & \cdots   & 0 & \frac{\Omega}{4}\sin\nu & 0 & \frac{b}{2}  & ~ \\
~ & ~ & ~ & ~ & ~ & ~ & ~ & ~ & ~ & ~ & ~ & ~ & \ddots \\
\end{bmatrix},
}
\end{multline}
where $N_{\Sigma/\Delta} = N_1 \pm N_2$. The perturbed part, on the other hand, is given by
\begin{equation}
    V'=
    \begin{bmatrix}
\ddots & ~ & ~ & ~ & ~ & ~ & ~ & ~ & ~ & ~ & ~ & ~ & ~ \\
~ & 0 & 1 & 0 & 0 & 0 & 0 & \cdots & 0 & 0 & 0 & 0 & ~\\
~ & 1 & 0 & 0 & 0 & 0 & 0 & \cdots & 0 & 0 & 0 & 0 & ~\\
~ & 0 & 0 & 0 & 1 & 0 & 0 & \cdots  & 0 & 0 & 0 & 0 & ~ \\
~ & 0 & 0 & 1 & 0 & 0 & 0 & \cdots & 0 & 0 & 0 & 0 & ~ \\
~ & 0 & 0 & 0 & 0 & 0 & 1 & \cdots & 0 & 0 & 0 & 0 & ~\\
~ & 0 & 0 & 0  & 0 & 1 & 0 &\cdots  & 0 & 0 & 0  & 0 & ~\\
~ & \vdots & \vdots & \vdots & \ddots & \vdots & \vdots & \vdots & \ddots & \vdots & \vdots & \vdots & ~ \\\
~ & 0 & 0 & 0 & 0 & 0 & 0  & \cdots & 0  & 1 & 0 & 0  & ~ \\
~ & 0 & 0 & 0 & 0 & 0 & 0 & \cdots & 1 & 0 & 0 & 0 & ~ \\
~ & 0 & 0 & 0 & 0 & 0 & 0 & \cdots  & 0 & 0 & 0 & 1  & ~\\
~ & 0 & 0 & 0 & 0 & 0 & 0 & \cdots   & 0 & 0 & 1 & 0  & ~ \\
~ & ~ & ~ & ~ & ~ & ~ & ~ & ~ & ~ & ~ & ~ & ~ & \ddots \\
\end{bmatrix}.
\end{equation}
The eigenvalues and eigenvectors of the matrix $H_{0,{\rm F}}$ can be solved in terms of Bessel functions $J_k(\mp\frac{\Omega\cos\nu}{2N_1\omega})$ and $J_k(\mp\frac{\Omega\sin\nu}{2N_2\omega})$~\cite{son2009floquet, Deng2015}. The appropriate transformation of the basis can be obtained by considering the eigenvalue problem
\begin{equation}
    \left(\mathcal{H}(t)-i\frac{\partial}{\partial t}\right)\phi(t)=\lambda \phi(t),
\end{equation}
where $\mathcal{H}(t)=-\frac{b}{2}+\frac{\Omega}{2}\cos\nu \cos{N_1\omega t}+\frac{\Omega}{2}\sin \nu\cos{N_2\omega t}$. The eigenvector for the trivial solution $\lambda=-\frac{b}{2}$ would be
\begin{equation}
    \phi(t)=\exp\left[-i\left(\frac{\Omega \cos\nu}{2N_1\omega}\sin{N_1\omega t}+\frac{\Omega\sin\nu}{2N_2\omega}\sin{N_2\omega t}\right)\right]=\sum_{k_1,k_2=-\infty}^{+\infty}J_{k_1}\left(-\frac{\Omega\cos\nu}{2N_1\omega}\right)J_{k_2}\left(-\frac{\Omega\sin\nu}{2N_2\omega}\right)e^{i(k_1N_1+k_2N_2)\omega t}.
\end{equation}
Similarly, for the choice of $\mathcal{H}(t)=\frac{b}{2}+\frac{\Omega}{2}\cos\nu \cos{N_1\omega t}+\frac{\Omega}{2}\sin\nu\cos{N_2\omega t}$ and $\lambda=\frac{b}{2}$, we have the solution
\begin{equation}
    \phi(t)=\exp\left[i\left(\frac{\Omega\cos\nu}{2N_1\omega}\sin{N_1\omega t}+\frac{\Omega \sin \nu}{2N_2\omega}\sin{N_2\omega t}\right)\right]=\sum_{k_1=-\infty}^{+\infty}\sum_{k_2=-\infty}^{+\infty} J_{k_1}\left(\frac{\Omega\cos\nu}{2N_1\omega}\right)J_{k_2}\left(\frac{\Omega\sin \nu}{2N_2\omega}\right)e^{i(k_1N_1+k_2N_2)\omega t}.
\end{equation}
Now the eigenvectors for the solutions $\lambda=\mp \frac{b}{2}+mN_1\omega+lN_2\omega$ are given by
\begin{equation}
    \phi^{ml}(t)=\sum_{k_1=-\infty}^{+\infty}\sum_{k_2=-\infty}^{+\infty} J_{k_1-m}\left(\mp\frac{\Omega\cos\nu}{2N_1\omega}\right)J_{k_2-l}\left(\mp\frac{\Omega\sin\nu}{2N_2\omega}\right)e^{i(k_1N_1+k_2N_2)\omega t}.
\end{equation}

To make our analysis of the Floquet matrix $H_{\rm F}$ convenient, we can enforce a change of basis of $H_{0,{\rm F}}$. The new basis in which $H_{0,{\rm F}}$ is diagonal is related to the existing one in the following way
\begin{equation}
    \ket{\phi^{ml}_\pm}=\sum_{k_1=-\infty}^{+\infty}\sum_{k_2=-\infty}^{+\infty} J_{k_1-m}\left(\mp\frac{\Omega\cos\nu}{2N_1\omega}\right)J_{k_1-l}\left(\mp\frac{\Omega\sin \nu}{2N_2\omega}\right)\ket{u_{\pm}^{k_1 k_2}}.
\label{eq:basis_change_onequbit}    
\end{equation}
The next step is to write the Floquet matrix $H_{\rm F}$ in the changed basis. We can do this because the difference between $H_{\rm F}$ and $H_{0,{\rm F}}$ is just a perturbative term. First, we calculate the off-diagonal elements of this new matrix. We start this exercise by making the following observation.
\begin{equation}
    \bra{u_{\pm}^{ml}}H_{\rm F}\ket{u_{\mp}^{jk}}=-\frac{w_{\rm q}}{2}\delta_{mj}\delta_{lk}.
\label{eq:off_diag}
\end{equation}
Using Eqs.~\eqref{eq:basis_change_onequbit} and \eqref{eq:off_diag}, and the following identities
\begin{equation}
    J_k(-x)=J_{-k}(x)\;\;\mathrm{and} \sum_{k=-\infty}^{+\infty}J_{n+k}(x)J_{n-k}(x)=J_n(2x),
\end{equation}
we obtain
\begin{equation}
    \bra{\phi_\mp^{ml}}H_F\ket{\phi_\pm^{jk}}=-\frac{w_{\rm q}}{2} J_{\pm({j-m})}(\frac{\Omega\cos \nu}{N_1\omega})J_{\pm({k-l})}(\frac{\Omega\sin \nu}{N_2\omega}).
\label{eq:new_off_diag}   
\end{equation}
Similarly, the diagonal entries of the new matrix can be obtained by using the fact
\begin{equation}
    \bra{u_{\pm}^{ml}}H_F\ket{u_{\pm}^{jk}}=(\mp \frac{b}{2}+mN_1\omega+lN_2\omega)\delta_{mj}\delta_{lk}\mp\frac{\Omega}{4}\sin \nu(\delta_{mj}\delta_{l,k+1}+\delta_{mj}\delta_{l,k-1})\mp\frac{\Omega}{4}\cos \nu(\delta_{m,j+1}\delta_{lk}+\delta_{m,j-1}\delta_{lk})
\label{eq:diag}
\end{equation}
and the identities
\begin{eqnarray}
    J_{k-1}(x)+J_{k+1}(x)=\frac{2k}{x}J_k(x)\nonumber\\
    \sum_k J_k(x)+J_{k-n}(x)=\sum_k J_k(x)+J_{n-k}(-x)=J_n(0)=\delta_{n0}.
\end{eqnarray}
Therefore
\begin{equation}
    \bra{\phi_\pm^{ml}}H_F\ket{\phi_\pm^{jk}}=(\mp \frac{b}{2}+mN_1\omega+lN_2\omega)\delta_{mj}\delta_{lk}.
\label{eq:new_diag}    
\end{equation}
Eq.~\eqref{eq:new_off_diag} and \eqref{eq:new_diag} jointly give the Floquet matrix in the new basis.

We define,
\begin{equation}
D_{m,l}^{0,0}=\begin{bmatrix}
\bra{\phi^{ml}_+}H_F\ket{\phi^{ml}_+} & \bra{\phi^{ml}_+}H_F\ket{\phi^{ml}_-}\\
\bra{\phi^{ml}_-}H_F\ket{\phi^{ml}_+} & \bra{\phi^{ml}_-}H_F\ket{\phi^{ml}_-}
\end{bmatrix}=
\begin{bmatrix}
-\frac{b}{2} + mN_1\omega + lN_2\omega & -\frac{w_{\rm q}}{2} J_0\left(\frac{\Omega\cos \nu}{N_1\omega}\right)J_0 \left(\frac{\Omega\sin\nu}{N_2\omega}\right) \\
-\frac{w_{\rm q}}{2} J_0\left(\frac{\Omega\cos\nu}{N_1\omega}\right)J_0 \left(\frac{\Omega\sin\nu}{N_2\omega}\right)  & \frac{b}{2} + mN_1\omega + lN_2\omega
\end{bmatrix},
\label{eq:Dml00}
\end{equation}
and for $\kappa_l,\kappa_m\neq 0$,
\begin{equation}
D_{m,l}^{\kappa_m, \kappa_l}=\begin{bmatrix}
0 & \bra{\phi^{ml}_+}H_F\ket{\phi^{m+\kappa_m\; l+\kappa_l}_-}\\
\bra{\phi^{ml}_-}H_F\ket{\phi^{m+\kappa_m \;l+\kappa_l}_+} & 0
\end{bmatrix}.
\end{equation}
In the new basis, the Floquet Hamiltonian $H_{\rm F}$ can be rewritten as
\begin{equation}
\tilde{H}_F =
\begin{bmatrix}
\ddots & ~ & ~ & ~ & \scalebox{-1}[1]{$\ddots$} \\
~ & \tilde{H}_F^{(m-1,0)}  & \tilde{H}_F^{(m-1,1)}  & \tilde{H}_F^{(m-1,2)}  ~\\
~ & \tilde{H}_F^{(m,-1)}  & \tilde{H}_F^{(m,0)}  & \tilde{H}_F^{(m,1)} & ~\\
~ & \tilde{H}_F^{(m+1,-2)}  & \tilde{H}_F^{(m+1,-1)}  & \tilde{H}_F^{(m+1,0)} & ~\\
\scalebox{-1}[1]{$\ddots$} & ~ & ~ & ~ & \ddots
\end{bmatrix},
\label{eq:floq_Ham_tran}
\end{equation}
where
\begin{equation}
\tilde{H}_F^{(m,\kappa_m)} =
\begin{bmatrix}
\ddots & ~ & ~ & ~ & \scalebox{-1}[1]{$\ddots$} \\
~ & D_{m,l-1}^{\kappa_m, 0} & D_{m,l-1}^{\kappa_m,1} & D_{m,l-1}^{\kappa_m,2} ~\\
~ & \left[{D_{m,l}^{\kappa_m,-1}}\right]^\dagger & D_{m,l}^{\kappa_m,0} & D_{m,l}^{\kappa_m,1} & ~\\
~ & \left[{D_{m,l+1}^{\kappa_m,-2}}\right]^\dagger & \left[{D_{m,l+1}^{\kappa_m,-1}}\right]^\dagger & D_{m,l+1}^{\kappa_m,0} & ~\\
\scalebox{-1}[1]{$\ddots$} & ~ & ~ & ~ & \ddots
\end{bmatrix}.
\end{equation}
We can break the above Hamiltonian into diagonal and off-diagonal elements and consider all the off-diagonal elements as perturbation. However, there are many off-diagonal elements and they will compete against each other. We will first try to separate out the most significant off-diagonal terms under different approximations and cancel out the terms with negligible contribution. Let's start with the Hamiltonian with only one off-diagonal element
\begin{equation}
\tilde{H}_F^{(0)} =
\begin{bmatrix}
\ddots & ~ & ~ & ~ & \scalebox{-1}[1]{$\ddots$} \\
~ & \tilde{H}_{F,0}^{(m-1,0)}  & 0  & 0  ~\\
~ & 0 & \tilde{H}_{F,0}^{(m,0)}  & 0 & ~\\
~ & 0  & 0  & \tilde{H}_{F,0}^{(m+1,0)} & ~\\
\scalebox{-1}[1]{$\ddots$} & ~ & ~ & ~ & \ddots
\end{bmatrix},
\end{equation}
where 
\begin{equation}
\tilde{H}_{F,0}^{(m,0)} =
\begin{bmatrix}
\ddots & ~ & ~ & ~ & \scalebox{-1}[1]{$\ddots$} \\
~ & D_{m,l-1}^{0, 0} &0 & 0 ~\\
~ & 0& D_{m,l}^{0,0} & 0 & ~\\
~ & 0& 0& D_{m,l+1}^{0,0} & ~\\
\scalebox{-1}[1]{$\ddots$} & ~ & ~ & ~ & \ddots
\end{bmatrix},
\end{equation}
with $D_{m,l}^{0,0}$ given by Eq.~(\ref{eq:Dml00}).

The Hamiltonian $\tilde{H}_F^{(0)}$ is a good approximation in the very strong driving regime where $\Omega\gg \omega$. In the strong driving regime, the two quasi-energies will be separated by $\frac{w_{\rm q}} {2}J_0\left(\frac{\Omega\cos\nu}{N_1\omega}\right)J_0 \left(\frac{\Omega\sin\nu}{N_2\omega}\right)$ which is the only off-diagonal element present in the Hamiltonian. However, when one of the drives becomes weaker, the higher order $k>0$ Bessel function becomes larger, and the contribution from other off-diagonal elements cannot be neglected. Next, we will study the quasienergies when one of the drives is strong whereas the other one is weak, i.e., $\nu\ll\pi/4$. Without loss of generality, we choose $m=l=0$. Following Ref.~\onlinecite{Deng2015}, the Floquet Hamiltonian takes the following form
\begin{equation}
\tilde{H}_{\rm F}^{(1)} = 
\begin{bmatrix}
~ & \tilde{H}_F^{(-1,0)}  & \tilde{H}_F^{(-1,1)}  ~\\
~ & \tilde{H}_F^{(0,-1)}  & \tilde{H}_F^{(0,0)}  
\end{bmatrix},
\end{equation}
where only $m,l = 0 {\rm~or~}-1$ contributions survive at resonance and all other contributions can be neglected. Further, we have
\begin{equation}
\tilde{H}_F^{(m,\kappa_m)} =
\begin{bmatrix}
~ & D_{m,-1}^{\kappa_m, 0} & D_{m,-1}^{\kappa_m,1}~\\
~ & \left[{D_{m,0}^{\kappa_m,-1}}\right]^\dagger & D_{m,0}^{\kappa_m,0} 
\end{bmatrix}.
\end{equation}
Expanding the Floquet matrix, and using the notation $\tilde{J}_{k,1}=J_k\left(\frac{\Omega\cos\nu}{N_1\omega}\right)$ and $\tilde{J}_{k,2}=J_k\left(\frac{\Omega\sin\nu}{N_2\omega}\right)$, we obtain
\begin{multline}
\tilde{H}_{\rm F}^{(1)} =\\
\begin{bmatrix}
-\frac{b}{2} -(N_1+N_2)\omega & -\frac{w_{\rm q}}{2} \tilde{J}_{0,1} \tilde{J}_{0,2} & 0 & 0 & 0 & \frac{w_{\rm q}}{2} \tilde{J}_{1,1}\tilde{J}_{0,2} & 0 & \frac{w_{\rm q}}{2} \tilde{J}_{1,1}\tilde{J}_{1,2} \\
-\frac{w_{\rm q}}{2} \tilde{J}_{0,1} \tilde{J}_{0,2} & \frac{b}{2} -(N_1+N_2)\omega  & 0 & 0 & -\frac{w_{\rm q}}{2} \tilde{J}_{1,1} \tilde{J}_{0,2}  & 0 & \frac{w_{\rm q}}{2} \tilde{J}_{1,1}\tilde{J}_{1,2}  & 0\\
0 & 0 & -\frac{b}{2} - N_1\omega & -\frac{w_{\rm q}}{2} \tilde{J}_{0,1} \tilde{J}_{0,2} & 0 & \frac{w_{\rm q}}{2} \tilde{J}_{1,1}\tilde{J}_{1,2}  & 0 & \frac{w_{\rm q}}{2} \tilde{J}_{1,1} \tilde{J}_{0,2} \\
0 & 0 & -\frac{w_{\rm q}}{2} \tilde{J}_{0,1} \tilde{J}_{0,2} & \frac{b}{2} - N_1\omega & \frac{w_{\rm q}}{2} \tilde{J}_{1,1}\tilde{J}_{1,2} & 0 & -\frac{w_{\rm q}}{2} \tilde{J}_{1,1} \tilde{J}_{0,2} & 0\\
0 & -\frac{w_{\rm q}}{2} \tilde{J}_{1,1} \tilde{J}_{0,2} & 0 & \frac{w_{\rm q}}{2} \tilde{J}_{1,1}\tilde{J}_{1,2} & -\frac{b}{2} - N_2\omega & -\frac{w_{\rm q}}{2} \tilde{J}_{0,1} \tilde{J}_{0,2} & 0 & 0 \\
\frac{w_{\rm q}}{2} \tilde{J}_{1,1} \tilde{J}_{0,2} & 0 & \frac{w_{\rm q}}{2} \tilde{J}_{1,1}\tilde{J}_{1,2} & 0 & -\frac{w_{\rm q}}{2} \tilde{J}_{0,1} \tilde{J}_{0,2} & \frac{b}{2} - N_2\omega & 0 & 0\\
0 & \frac{w_{\rm q}}{2} \tilde{J}_{1,1}\tilde{J}_{1,2} & 0 & -\frac{w_{\rm q}}{2} \tilde{J}_{1,1} \tilde{J}_{0,2} & 0 & 0 & -
\frac{b}{2}  & -\frac{w_{\rm q}}{2} \tilde{J}_{0,1} \tilde{J}_{0,2}
\\
\frac{w_{\rm q}}{2} \tilde{J}_{1,1}\tilde{J}_{1,2} & 0 & \frac{w_{\rm q}}{2} \tilde{J}_{1,1} \tilde{J}_{0,2} & 0 & 0 & 0 & -\frac{w_{\rm q}}{2} \tilde{J}_{0,1} \tilde{J}_{0,2} & \frac{b}{2}
\end{bmatrix}.
\end{multline}
In the fast driving regime $(\omega \approx w_{\rm q})$, the quasienergies are given by the following Hamiltonian
\begin{equation}
\tilde{H}_{\rm F}^{(1)} =\begin{bmatrix}
-\frac{b}{2} - (N_1+N_2)\omega & -\frac{w_{\rm q}}{2}\tilde{J}_{0,1} \tilde{J}_{0,2}  & 0 & \frac{w_{\rm q}}{2}\tilde{J}_{1,1}\tilde{J}_{1,2} \\
-\frac{w_{\rm q}}{2} \tilde{J}_{0,1} \tilde{J}_{0,2} & \frac{b}{2} -(N_1+N_2)\omega   & \frac{w_{\rm q}}{2} \tilde{J}_{1,1} \tilde{J}_{1,2}  & 0\\

0 & \frac{w_{\rm q}}{2} \tilde{J}_{1,1} \tilde{J}_{1,2}  & -\frac{b}{2} - N_2\omega & -\frac{w_{\rm q}}{2} \tilde{J}_{0,1} \tilde{J}_{0,2} \\
\frac{w_{\rm q}}{2} \tilde{J}_{1,1} \tilde{J}_{1,2}  & 0 & -\frac{w_{\rm q}}{2} \tilde{J}_{0,1} \tilde{J}_{1,2} & \frac{b}{2} - N_2\omega \\
\end{bmatrix}.
\end{equation}
Substituting, $(N_1+N_2)\omega\to \omega$, we obtain the Floquet quasienergy gap given by
\begin{equation}
\Delta\epsilon={\omega}-\bigg\{\Theta^2 + \omega^2 + w_{\rm q}^2  \tilde{J}_{1,1}^2\tilde{J}_{1,2}^2- 2\Theta \omega\bigg\}^{1/2},
\end{equation}
where we defined $\Theta^2 = b^2 + w_{\rm q}^2 \tilde{J}_{0,1}^2\tilde{J}_{0,2}^2 $. The above expression for the quasienergy gap agrees well with numerical results in the fast-driving regime, particularly when one drive is much weaker than the other ($\nu\ll \pi/4$). Notably, qualitative agreement persists even beyond this regime.

\subsection{Multiphoton Resonance: AC Stark Shift and Power Broadening}
\label{sec:lev_shift}
Using Eq.~(\ref{eq:floq_Ham_tran}), the generalised Van-Vleck (GVV) Hamiltonian for the degeneracy condition $-b/2 \approx b/2 - mN_1\omega - lN_2\omega$ is given by~\cite{son2009floquet},
\begin{equation}
         H_{\rm GVV}=
            \begin{pmatrix}
             -b/2 +\chi & -({w_{\rm q}}/{2}) \tilde{J}_{-m,1}\tilde{J}_{-l,2}\\
            -({w_{\rm q}}/{2}) \tilde{J}_{-m,1}\tilde{J}_{-l,2} & b/2 -mN_1\omega - lN_2 \omega -\chi
            \end{pmatrix}.
\label{eq:app_HGVV}            
\end{equation}
To calculate the level shifts $\chi$ corresponding to the transition $\ket{\phi^+_{00}}$ to $\ket{\phi^-_{-m,-l}}$, we apply the nearly degenerate GVV perturbation method~\cite{son2009floquet}. The $2\times 2$ matrix $\mathcal{H}$ and its eigenstates $\phi$ can be expanded in powers of $-w_{\rm q}/2$ as follows
\begin{equation}
    \mathcal{H}=\sum_{k=0}^\infty (-w_{\rm q}/2)^k \mathcal{H}^{(k)} \;\; {\rm and} \;\; \phi=\sum_{k=0}^\infty (-w_{\rm q}/2)^k \phi^{(k)}.
\end{equation}

Assuming that the states $\ket{\phi^+_{00}}$ to $\ket{\phi^-_{-m,-l}}$ are nearly degenerate we have $-b/2\approx b/2 -mN_1\omega-lN_2\omega$. Let the zeroth order state $\ket{\phi^{(0)}}=\{\phi^{(0)}_+ , \phi^{(0)}_-\}$, such that $\phi^{(0)}_+=\ket{\phi_+^{00}}$ and $\phi^{(0)}_-=\ket{\phi_-^{-m,-l}}$.
The zeroth order correction in $\mathcal{H}$ is then given by
\begin{equation}
    \mathcal{H}^{(0)}=\begin{pmatrix}
                        -b/2 && 0\\
                        0 && b/2 -mN_1\omega-lN_2\omega
                       \end{pmatrix} 
\end{equation}

Now calculating the first order terms in the expansion of $\mathcal{H}$ using the GVV method~\cite{HIRSCHFELDER19781, son2009floquet}, we have
\begin{equation}
    \mathcal{H}^{(1)}= \bra{\phi^{(0)}}V'\ket{\phi^{(0)}}=\tilde{J}_{-m,1}\tilde{J}_{-l,2}\begin{pmatrix}
                     0 && 1\\
                     1 && 0
                    \end{pmatrix}.
 \label{eq:order1_onequbit}            
\end{equation}
Further using the techniques in Ref~\cite{HIRSCHFELDER19781, son2009floquet}, we obtain
\begin{eqnarray}
    \phi^{(1)}_+ &=& \sum_{\substack{j,k=-\infty\\j,k\neq -m,-l}}^{+\infty} \frac{-\tilde{J}_{j,1}\tilde{J}_{k,2}}{(b +jN_1\omega+kN_2\omega)}\ket{\phi^-_{jk}},\nonumber\\
    \phi^{(1)}_- &=& \sum_{\substack{j,k=-\infty\\j,k\neq -m,-l}}^{+\infty} \frac{\tilde{J}_{j,1}\tilde{J}_{k,2}}{(b +jN_1\omega+kN_2\omega)}\ket{\phi^+_{-j-m,-k-l}},
\end{eqnarray}
using which we can compute the next-order corrections in $\mathcal{H}$ given by
\begin{equation}
    \mathcal{H}^{(2)} = \bra{\phi^{(0)}}V'\ket{\phi^{(1)}}-\mathcal{H}^{(1)}\langle{\phi^{(0)}}|\phi^{(1)}\rangle=\sum_{\substack{j,k=-\infty\\j,k\neq -m,-l}}^{+\infty} \frac{\tilde{J}_{j,1}^2\tilde{J}_{k,2}^2}{(b +jN_1\omega+kN_2\omega)}\begin{pmatrix}
                       -1 & 0\\
                        0 & 1
                    \end{pmatrix}.
 \label{eq:order2_onequbit}
 \end{equation}
Now comparing the matrix structures in Eq.~\eqref{eq:order1_onequbit} and \eqref{eq:order2_onequbit} with Eq.\eqref{eq:app_HGVV}, we see that  $\chi$ can be related to the odd powers of $w_{\rm q}/\omega$,
\begin{equation}
    \chi=-\frac{w_{\rm q}^2}{4}\sum_{\substack{j,k=-\infty\\j,k\neq -m,-l}}^{+\infty} \frac{\left(\tilde{J}_{j,1}\right)^2\left(\tilde{J}_{k,2}\right)^2}{(b +jN_1\omega+kN_2\omega)} + \mathcal{O}\left(\frac{w_{\rm q}}{\omega}\right)^3.
\label{eq:Stark_Shift_onequbit}
\end{equation}

\end{widetext}

\bibliography{references}
\end{document}